\documentclass[preprint,prl]{revtex4-1}
\usepackage{epsf}
\usepackage{amsmath,amsthm,amssymb}
\usepackage{color}
\usepackage{dcolumn}
\usepackage{bm}
\usepackage{graphicx,epsfig}

\graphicspath{{fig/}}
\everymath{\displaystyle}

\begin{document}           

\title{Rotation of intrinsic orbital angular momentum and the orbital Hall effect for twisted particles in arbitrary gravitational fields}

\author{Alexander J. Silenko} \email{alsilenko@mail.ru}
\affiliation{Bogoliubov Laboratory of Theoretical Physics, Joint Institute for Nuclear Research, Dubna 141980, Russia}
\affiliation{Moscow Center for Advanced Studies, 
Kulakova str. 20, Moscow 123592, Russia}

\date{\today}

\begin {abstract}
For twisted particles in arbitrary gravitational fields, the problems of the rotation of intrinsic orbital angular momentum and the orbital Hall effect are solved in the general case. 
We need not use the Maxwell equations in curved spacetimes for a description of twisted photons.
The exact general equation rigorously defining the OAM dynamics in any Riemannian spacetimes is derived. The general description of different manifestations of the orbital Hall effect for any twisted particles is also presented. Our short analysis shows that the results obtained can find important practical applications.
\end{abstract}


\maketitle


Nowadays, twisted (vortex) particles possessing intrinsic orbital angular momenta (OAMs) take an important place in fundamental and applied physics \cite{BliokhSOI,Lloyd,LarocqueTwEl,Ivanov:2022jzh,Innovation}. It has been argued in Ref. \cite{Katoh} that twisted photons are very ubiquitous in laboratories and in nature. Next studies confirmed this conclusion and extended it to other particles (see, e.g., Refs. \cite{Harwit,Tambu,ICHEP2023}). Light can acquire an OAM when it traverses the gravitational field of a massive rotating compact object \cite{Tamburini}. The use of twisted photons in free-space optical communications, including information transfer, is very promising \cite{twistfreespace,twistfreesptwo,twistrevie,twistnanop,photonIEEE,photonCSp,photonnew}. An observation of twisted light allows one to determine the spin of a black hole \cite{Tamburi}. However, a detailed analysis of the problem requires a detailed theoretical description of OAM dynamics in gravitational fields and noninertial frames. This analysis was not previously performed. We can mention only investigations of the trajectory of a twisted particle in gravitational fields \cite{photontwistd} and the orbital Hall effect of vortex light \cite{orbitalHalleff}.

In this Letter, we rigorously derive a general and exact classical equation describing dynamics of the OAM in arbitrary Riemannian spacetimes. The equation is valid in the framework of general relativity and can find important practical applications. We also present a general description of the orbital Hall effect for any twisted particles in arbitrary Riemannian spacetimes and detect a violation of the weak equivalence principle (WEP). Amazingly, we need no Maxwell equations in curved spacetimes for a description of twisted photons.

In the present work, we use the system of units $\hbar=1,~c=1$ but explicitly include $\hbar$ and $c$ when this inclusion clarifies the problem. We denote world and spatial indices by Greek and Latin letters
$\alpha,\mu,\nu,\ldots$ $=0,1,2,3,~i,j,k,\ldots=1,2,3$, respectively. Tetrad
indices are denoted by Latin letters from the beginning of the
alphabet, $a,b,c,\ldots = 0,1,2,3$. Temporal and spatial tetrad indices are
distinguished by hats. The signature is $(+---)$.
Commas and semicolons before indices denote partial and covariant derivatives, respectively.

The equation of motion for the orbital polarization of twisted particles in electrodynamics has been obtained in Refs. \cite{Manipulating,ResonanceTwistedElectrons} on the basis of the Lorentz transformations for electric and magnetic dipole moments (EDMs and MDMs) derived in Ref. \cite{PhysScr}. 
The equation for the OAM precession in electrodynamics reads \cite{Manipulating,ResonanceTwistedElectrons}
 \begin{equation}\begin{array}{c}
\frac{d\bm L}{dt}=\bm\Omega\times\bm L,\qquad \bm\Omega=-\frac{e}{2m\gamma}\left(\bm B-\bm\beta\times\bm E\right).
\end{array} \label{reLarpn} \end{equation}

This equation substantionally differs from the Thomas-Bargmann-Mishel-Telegdi equation for the spin precession \cite{Thomas,BMT} (see also Ref. \cite{PhysScr}). However, the conventional three-component spin is defined in the particle rest frame and the angular velocity of its precession includes the correction for the Thomas effect. In contrast, the OAM is defined in the laboratory frame. 

Nevertheless, the OAM and spin are two forms of the same physical quantity, angular momentum. These forms satisfy the same transformation relations and their substantional difference follows from fundamental properties of the Poincar\'{e} group. Specifically, the total angular momentum is the sum of the orbital and spin parts, $\bm j=\bm L+\bm s$, and only the definition of the OAM and spin in the laboratory and particle rest frames, respectively, allows one to use the generally accepted commutative geometry (see Ref. \cite{PRAFW}). Alternative definitions are also possible but they lead to noncommutative geometries \cite{PRAFW}. In intermediary calculations, the OAM and spin can be defined in any frames.

In gravity, it is convenient to apply local Lorentz frames (coframes, tetrad frames). Local Lorentz frames (LLFs) are characterized by the Minkowski metric $ds^2=\eta_{ab}dx^adx^b$, where $\eta_{ab}={\rm diag}(1, -1, -1, -1)$. The metric tensor of a given spacetime can be split into tetrads $e^a_\mu$. Any tetrad can be attributed to an observer. Observers carrying different tetrads move relative to each other with a definite velocity $\bm V$ \cite{PRDThomaspre}. The connection between two LLFs attributed to these observers is realized by local Lorentz transformations, and $\bm V$ can depend on time \cite{PRDThomaspre}. Although any tetrad is applicable, the preferable choice is the Schwinger tetrad \cite{Schwinger,dirac} carried by an observer at rest in a given spacetime (see Refs. \cite{PRDThomaspre,ost,gravityDirac,gravityDiracOST1,gravityDiracOST2,gravityDiracOST3,gravityDiracOST4,gravityDiracOST5} for more details). This tetrad is defined by $e_i^{\hat{0}}=0,~e_{\hat{i}}^{0}=0$ $(i=1,2,3)$ and specifies the laboratory frame. Such a choice has been used for derivations of quantum-mechanical and classical equations of spin motion in arbitrary Riemannian \cite{PRDThomaspre,ost,gravityDirac,gravityDiracOST1,gravityDiracOST2,gravityDiracOST4,gravityDiracOST5} and Riemann-Cartan \cite{gravityDiracOST3} spacetimes.
General classical and quantum-mechanical Hamiltonians for a particle (or a test body) in Riemannian spacetimes have been derived in Refs. \cite{Cogn,gravityDiracOST2} and \cite{PRDThomaspre,ost,gravityDirac,gravityDiracOST1,gravityDiracOST2,gravityDiracOST4,gravityDiracOST5,otherspin1,otherspin2}, respectively.


A particle carrying an \emph{intrinsic} OAM can be described by a LG beam. A beam produced in a laboratory can be modeled by a moving centroid which is formed by a continuum of partial de Broglie waves. The centroid has the axis of symmetry $z$ and the orthogonal plane of symmetry containing the beam waist \cite{BliokhSOI,Lloyd}. The average radius and kinetic momentum of the beam vanish in the transverse plane ($\langle\boldsymbol{\mathfrak{r}}\rangle=0$, $\bm{p}_\bot=0$, $\langle\boldsymbol{\mathfrak{p}}_\bot\rangle=0$). The canonical momentum of a specific partial wave is equal to $\bm p+\boldsymbol{\mathfrak{p}}$, where $\bm p$ is the beam momentum and $\boldsymbol{\mathfrak{p}}$ defines the additional motion of this wave. In Riemannian spacetimes, we introduce these parameters in LLFs and distinguish them by hats. 

Our denotation for the intrinsic OAM defined in the laboratory coframe is $\widehat{\bm L}=\widehat{\boldsymbol{\mathfrak{r}}}\times\widehat{\boldsymbol{\mathfrak{p}}}$. For an observer in the coframe instantaneously accompanying the centroid (CIAC), it takes the form $\widehat{\bm L}^{(0)}=\widehat{\boldsymbol{\mathfrak{r}}}^{(0)}\times\widehat{\boldsymbol{\mathfrak{p}}}^{(0)}$. In Minkowski space, averaging over partial waves results in $\langle(\bm p+\boldsymbol{\mathfrak{p}})^2\rangle=\bm p^2+\langle\boldsymbol{\mathfrak{p}}_\bot^2\rangle$ and the centroid has \emph{effective} mass $M=\sqrt{m^2+\langle\boldsymbol{\mathfrak{p}}_\bot^2\rangle}$ \cite{photonPRA}. For photons, $M=\sqrt{\langle\boldsymbol{\mathfrak{p}}_\bot^2\rangle}$. Nonzero values of $M$ lead to \emph{subluminal} velocities \cite{Giovannini,AlfanoNolanBessel,Bouchard,Lyons} of twisted light. The intrinsic OAM is equal to $\widehat{\boldsymbol{\mathcal{L}}}$. After a Lorentz boost, the intrinsic OAM can take \emph{any} orientation relative to the centroid momentum. Therefore, the often used relation (see, e.g., the review \cite{MOancea}) 
\begin{equation}
\bm L=\hbar\ell\frac{\bm p}{p}
\label{ellheli} \end{equation} is invalid. However, it can be shown (see Supplemental Material for details) that this relation usually leads to \emph{approximately} correct results because photon centroids are ultrarelativistic. However, a photon centroid being effectively massive is characterized by the rest frame and $2\ell+1$ OAM projections on a chosen direction. In contrast, Eq. (\ref{ellheli}) predicts that the OAM has only two projections $\pm L$ on the momentum direction. This prediction is mistaken, although the photon spin has two projections on the momentum direction \emph{for any partial plane wave}.

We describe the OAM dynamics on the basis of gravitoelectromagnetic fields introduced by Pomeransky and Khriplovich \cite{PK} (see also Ref. \cite{Obzor}) and then investigated in Refs. \cite{gravityDirac,gravityDiracOST1,gravityDiracOST2,gravityDiracOST3,gravityDiracOST4,gravityDiracOST5,PRDThomaspre} in detail. Our explanations mainly follow Ref. \cite{PRDThomaspre}.

The equation of motion of a pointlike spinless particle in general relativity reads
\begin{equation}
\begin{array}{c}
Du^{\mu}= 0.
\end{array} \label{eqPKu} \end{equation}
This equation defines the particle motion on a geodesic line which is perturbed by the Mathisson force for spinning particles and by tidal forces for extended ones. As a rule, the Mathisson and tidal forces are relatively small and can be neglected in the present study. 
The orthogonality condition interconnects the four-vectors of velocity $u^\mu$ and angular momentum $a^\mu$ if we define the latter quantity in the centroid rest frame. Since we consider LLFs, $a^{a}=(0,\widehat{\boldsymbol{\mathcal{L}}})$ and
\begin{equation}
\begin{array}{c}
u^{\mu}a_\mu=u^{a}a_a= 0.
\end{array} \label{eqMatPi} \end{equation}
Equations (\ref{eqPKu}) and (\ref{eqMatPi}) result in (see Refs. \cite{gravityDirac,PRDThomaspre})
\begin{equation}
\begin{array}{c}
Da^{\mu}= 0,\qquad \frac{Du^{a}}{d\tau}=\frac{du^{a}}{d\tau}=\Gamma^{a}_{~bc} u^b u^c,\\ \frac{Da^{a}}{d\tau}=\frac{da^{a}}{d\tau}= \Gamma^a_{~bc}a^bu^c.
\end{array} \label{eqPKn} \end{equation}
Here $\Gamma_{abc}=-\Gamma_{bac}= e_b^\mu e_c^\nu e_{a\mu;\nu}$ are the Lorentz connection coefficients (Ricci rotation coefficients). They can also be presented in the form
\begin{equation}\begin{array}{c}
\Gamma_{abc}=\frac 12\left(\lambda_{abc}+\lambda_{bca}-\lambda_{cab}\right),\\
\lambda_{abc}=-\lambda_{acb}=e_b^\mu e_c^\nu (e_{a\mu,\nu}-e_{a\nu,\mu}).
\end{array}\label{eqin7}\end{equation} 

Equation (\ref{eqPKn}) is similar to the corresponding equations in electrodynamics for the four-momentum and four-spin of a particle without an anomalous magnetic moment,
\begin{equation}
\begin{array}{c}
\frac{du_{\mu}}{d\tau}=\frac{e}{mc}F_{\mu\nu} u^\nu,\qquad
\frac{ds_{\mu}}{d\tau}=\frac{e}{mc}F_{\mu\nu} s^\nu.
\end{array} \label{eqThBMT} \end{equation}
A deep similarity between $(e/m)F_{\mu\nu}$ and $\Gamma_{abc}u^c$ has allowed Pomeransky and Khriplovich to introduce the gravitoelectric and gravitomagnetic fields $c\Gamma_{abc}u^c=\mathcal{F}_{ab}=(\bm{\mathcal{E}},\bm{\mathcal{B}})$. Explicitly \cite{PK}
\begin{equation}
\begin{array}{c} \mathcal{E}_{\widehat{i}}=c\Gamma_{0ic}u^{c},\qquad
\mathcal{B}_{\widehat{i}}=-\frac{c}{2}e_{ikl}\Gamma_{klc}u^{c}.
\end{array} \label{expl}
\end{equation} We do not make a difference between upper and lower indices of these fields.

The equations of motion for the centroid four-velocity read (see Refs. \cite{PK,PRDThomaspre})
\begin{equation}
\begin{array}{c} \frac{d\widehat{\bm{u}}}{d\tau}=u^{\widehat{0}}\bm{\mathcal{E}}+
\widehat{{\bm
u}}\times\bm{\mathcal{B}},  \qquad \frac{d{u}^{\widehat{0}}}{d\tau}=
\bm{\mathcal{E}}\cdot\widehat{\bm u}.
\end{array} \label{force}
\end{equation}
The particle acceleration in the coframe is given by \cite{PK,gravityDiracOST2,PRDThomaspre}
\begin{eqnarray}
\widehat{\bm w}\equiv\frac{d\widehat{\bm v}}{d\widehat{t}}=\frac {c}{{u}^{\widehat{0}}}\left[\frac{d\widehat{\bm u}}{d\widehat{t}}-\frac{\widehat{\bm u}}{\left({u}^{\widehat{0}}\right)^2}\left(\widehat{\bm u}\cdot \frac{d\widehat{\bm u}}{d\widehat{t}}\right)\right]
\nonumber\\=\frac {c}{{u}^{\widehat{0}}}\left[\bm{\mathcal{E}} 
+\widehat{\bm{\beta}} \times\bm{\mathcal{B}}-\widehat{\bm{\beta}} \left(\bm{\mathcal{E}}\cdot\widehat{\bm{\beta}}\right)\right],\label{RotFFor}
\end{eqnarray} where $\widehat{\bm{\beta}}=\widehat{\bm v}/c=\widehat{\bm u}/u^{\widehat{0}}=\widehat{\bm p}/\epsilon$, $\widehat{\bm p}=m\widehat{\bm u}$, and $\epsilon=\sqrt{M^2+\widehat{\bm p}^2}$.
We should add that $dt=u^0d\tau, \, d\widehat{t}=u^{\widehat{0}}d\tau=(u^{\widehat{0}}/u^0)dt$.

The angular momentum defined in the centroid rest frame becomes similar to the spin. 
Therefore, the equation of motion for the angular momentum defined in this frame is the same as the equation of spin motion \cite{PK,gravityDiracOST2,PRDThomaspre} and, therefore, reads
\begin{equation}
\begin{array}{c} \frac{d\hat{\boldsymbol{\mathcal{L}}}}{dt}=\bm{\Omega}^{(r)}\times \hat{\boldsymbol{\mathcal{L}}}, \qquad \bm{\Omega}^{(r)}=\frac{1}{u^0}\left(-\bm{\mathcal{B}}+
\frac{\widehat{\bm u}\times\bm{\mathcal{E}}}
{u^{\widehat{0}}+1}\right),
\end{array} \label{omgem}
\end{equation} where $u^0$ is defined in the world frame. For the intrinsic OAM, we should pass to the laboratory coframe quantity $\hat{\bm L}$. Hereinafter, the particle motion and gravitoelectromagnetic fields are defined relative to the laboratory coframe. It has been explicitly shown in electrodynamics \cite{PhysScr,Thomas,Jackson,DraganThomas} and gravity \cite{PRDThomaspre} that the difference between the spin motion in the instantaneously accompanying frame and the particle rest frame is defined by the Thomas precession. In the considered case \cite{PRDThomaspre}
\begin{equation}
\begin{array}{c} \bm{\Omega}^{(r)}=\bm{\Omega}^{(0)}+\bm{\Omega}^{(T)},\qquad \bm{\Omega}^{(T)}=-
\frac{u^{\widehat{0}}}{u^{\widehat{0}}+1}\left({\widehat{\bm\beta}}\times\frac{d\widehat{\bm\beta}}{dt}\right)\\=-
\frac{1}{u^{\widehat{0}}(u^{\widehat{0}}+1)}\left(\widehat{\bm{u}}\times\frac{d\widehat{\bm{u}}}{dt}\right).
\end{array} \label{omTom}
\end{equation}
It is convenient to pass to the laboratory coframe. The use of Eq. (\ref{RotFFor}) results in \cite{PRDThomaspre}
\begin{equation}\begin{array}{c}
\widehat{\bm{\Omega}}^{(T)}
=\frac{u^{\widehat{0}}-1}{u^{\widehat{0}}}\bm{\mathcal{B}}-\frac{1}{u^{\widehat{0}}(u^{\widehat{0}}+1)}\widehat{\bm u}(\widehat{\bm u}\cdot\bm{\mathcal{B}})-
\frac{1}
{u^{\widehat{0}}+1}\widehat{\bm u}\times\bm{\mathcal{E}},\\
\widehat{\bm{\Omega}}^{(0)}=-\bm{\mathcal{B}}+\frac{1}{u^{\widehat{0}}(u^{\widehat{0}}+1)}\widehat{\bm u}(\widehat{\bm u}\cdot\bm{\mathcal{B}})+
\frac{\widehat{\bm u}\times\bm{\mathcal{E}}}
{u^{\widehat{0}}}.
\end{array} \label{Thpexpl} \end{equation}
All angular velocities are given in the laboratory coframe and
\begin{equation}\begin{array}{c}
\Omega=\frac{d\widehat{\phi}}{dt},\qquad \widehat{\Omega}=\frac{d\widehat{\phi}}{d\widehat{t}},\qquad 
\widehat{\bm{\Omega}}=\frac{u^{0}}{u^{\widehat{0}}}\bm\Omega,\qquad \widehat{H}=\frac{u^{0}}{u^{\widehat{0}}}H.
\end{array} \label{conne} \end{equation}

The interaction Hamiltonian in this coframe reads
\begin{equation}
\widehat{H}=\widehat{\bm{\Omega}}\cdot\widehat{\bm L}.
\label{Thiaf} \end{equation}

The gravitoelectric and gravitomagnetic fields 
transform like the electric and magnetic ones \cite{PRDThomaspre}:
\begin{equation}\begin{array} {c}
\bm{\mathcal{E}}^{(0)}=u^{\widehat{0}}\left[\bm{\mathcal{E}}-\frac{u^{\widehat{0}}}{u^{\widehat{0}}+1}\widehat{\bm\beta}(\widehat{\bm\beta}\cdot\bm{\mathcal{E}})
+\widehat{\bm\beta}\times\bm{\mathcal{B}}\right],\\
\bm{\mathcal{B}}^{(0)}=u^{\widehat{0}}\left[\bm{\mathcal{B}}-\frac{u^{\widehat{0}}}{u^{\widehat{0}}+1}\widehat{\bm\beta}(\widehat{\bm\beta}\cdot\bm{\mathcal{B}})
-\widehat{\bm\beta}\times\bm{\mathcal{E}}\right].
\end{array} \label{meffinal} \end{equation}
Equations (\ref{Thpexpl}) --  (\ref{meffinal}) show that 
\begin{equation}\begin{array}{c}
\widehat{\bm{\Omega}}^{(0)}=-\frac{\bm{\mathcal{B}}^{(0)}}{u^{\widehat{0}}}.
\end{array} \label{omgfm}
\end{equation}
Equation (\ref{omgfm}) demonstrates that the precession of $\widehat{\bm L}^{(0)}$ in the \emph{laboratory} coframe does not depend on the gravitoelectric field in the CIAC. This conclusion remains valid for the precession of $\widehat{\bm L}^{(0)}$ in the CIAC. However, in this coframe 
\begin{equation}\begin{array}{c}
(u^{\widehat{0}})^{(0)}=1, \qquad (\widehat{\bm{\Omega}}^{(0)})^{(0)}=-\bm{\mathcal{B}}^{(0)}, \\
 \widehat{H}^{(0)}=(\widehat{\bm{\Omega}}^{(0)})^{(0)}\cdot\widehat{\bm L}^{(0)}=-\bm{\mathcal{B}}^{(0)}\cdot\widehat{\bm L}^{(0)}.
\end{array}\label{omiaf}
\end{equation}

Similarly to electrodynamics \cite{PhysScr}, it is convenient to introduce the gravitoelectric and gravitomagnetic dipole moments, $\widehat{\bm{\mathcal{D}}}$ and $\widehat{\bm{\mathcal{M}}}$:
\begin{equation}
\widehat{H}=-\widehat{\bm{\mathcal{D}}}\cdot\bm{\mathcal{E}}-\widehat{\bm{\mathcal{M}}}\cdot\bm{\mathcal{B}}.
\label{gemdm} \end{equation}
In the CIAC, $\widehat{\bm{\mathcal{D}}}^{(0)}=0$, $\bm{\mathcal{B}}=\bm{\mathcal{B}}^{(0)}$ and $\widehat{\bm{\mathcal{M}}}^{(0)}=\widehat{\bm L}^{(0)}$. In the laboratory coframe, $\widehat{\bm{\mathcal{D}}}$ is nonzero. Equation (\ref{gemdm}) for the gravitoelectromagnetic moments are similar to that for electromagnetic moments \cite{PhysScr}. The quantities $\widehat{\bm{\mathcal{D}}}^{(0)}$ and $\widehat{\bm{\mathcal{M}}}^{(0)}$ cannot form a four-tensor (more exactly, a four-tensor-like quantity) because the convolution of such tensor with $\mathcal{F}_{ab}$ should be an invariant. Like in electrodynamics \cite{PhysScr}, the antisymmetric four-tensor of gravitoelectromagnetic moments has the form
\begin{equation}
\mathcal{D}^{ab}=(u^{\widehat{0}}\widehat{\bm{\mathcal{D}}},u^{\widehat{0}}\widehat{\bm{\mathcal{M}}})=L^{ab},
\label{amten} \end{equation} where $L^{ab}$ is an antisymmetric coframe tensor of angular momentum. Since $(u^{\widehat{0}})^{(0)}=1$, the transformation of the gravitoelectromagnetic moments reads (cf. Refs. \cite{Manipulating,PhysScr})
\begin{equation}
\widehat{\bm{\mathcal{D}}}=\widehat{\bm\beta}\times\widehat{\bm{\mathcal{M}}}^{(0)},\qquad \widehat{\bm{\mathcal{M}}}=\widehat{\bm{\mathcal{M}}}^{(0)}\!-\frac{u^{\widehat{0}}}{u^{\widehat{0}}+1}\widehat{\bm\beta}(\widehat{\bm\beta}\cdot\widehat{\bm{\mathcal{M}}}^{(0)}).
\label{deqfinal} \end{equation}
The laboratory coframe Hamiltonian reads 
\begin{equation}\begin{array}{c}
\widehat{H}=\widehat{\bm{\Omega}}^{(0)}\cdot\widehat{\bm L}^{(0)}=-\biggl[\bm{\mathcal{B}}\cdot\widehat{\bm L}^{(0)}\\-\frac{u^{\widehat{0}}}{u^{\widehat{0}}+1}(\widehat{\bm\beta}\cdot\bm{\mathcal{B}})(\widehat{\bm\beta}\cdot\widehat{\bm L}^{(0)})-(\widehat{\bm\beta}\times\bm{\mathcal{E}})\cdot\widehat{\bm L}^{(0)}\biggr].
\end{array} \label{geiaf} \end{equation}
The equation
\begin{equation}\begin{array}{c}
\widehat{\bm L}=u^{\widehat{0}}\left[\widehat{\bm L}^{(0)}-\frac{u^{\widehat{0}}}{u^{\widehat{0}}+1}\widehat{\bm\beta}(\widehat{\bm\beta}\cdot\widehat{\bm L}^{(0)})\right]
\end{array} \label{oamopem} \end{equation} and Eq. (\ref{conne}) result in (cf. Ref. \cite{Manipulating})
\begin{equation}\begin{array}{c}
\widehat{H}=-\frac{1}{u^{\widehat{0}}}\left[\bm{\mathcal{B}}\cdot\widehat{\bm L}-(\widehat{\bm\beta}\times\bm{\mathcal{E}})\cdot\widehat{\bm L}\right],\\ H=\bm{\Omega}\cdot\widehat{\bm L}=-\frac{1}{u^{0}}\left[\bm{\mathcal{B}}\cdot\widehat{\bm L}-(\widehat{\bm\beta}\times\bm{\mathcal{E}})\cdot\widehat{\bm L}\right].
\end{array} \label{gelab} \end{equation}
The observable angular velocity of OAM rotation is given by
\begin{equation}
\frac{d\widehat{\bm L}}{dt}=\bm{\Omega}\times\widehat{\bm L},\qquad \bm{\Omega}=-\frac{1}{u^{0}}\left(\bm{\mathcal{B}}-\widehat{\bm\beta}\times\bm{\mathcal{E}}\right).
\label{final} \end{equation}

This general equation exhaustively describes the OAM precession in arbitrary Riemannian spacetimes. In Refs. \cite{gravityDiracOST2,gravityDiracOST3}, the exact formulas for the fields $\bm{\mathcal{E}}$ and $\bm{\mathcal{B}}$ have been derived in the general case. In some important cases, explicit expressions for these fields have been obtained in Ref. \cite{PRDThomaspre}. The use of the weak-field approximation in the Earth frame leads to \cite{PRDThomaspre}
\begin{equation}\begin{array}{c}
\bm{\mathcal{E}}=\frac{\bm gu^{\widehat{0}}}{c}, \qquad \bm{\mathcal{B}}=\bm\omega u^{\widehat{0}}+\frac{\bm g\times\widehat{\bm u}}{c},
\end{array}\label{gemCSch}\end{equation}
where $g=9.81$ m/s$^2$ is the Newtonian gravitational constant and $\omega=7.29\times10^{-5}$ s$^{-1}$ is the angular frequency of the Earth rotation. As follows from Eqs. (\ref{final}) and (\ref{gemCSch}), the OAM rotation is rather slow and can be neglected at the information transfer. In particular, the OAM rotates with the angular velocity \begin{equation}\bm{\Omega}_g=\frac{2u^{\widehat{0}}(\widehat{\bm\beta}\times\bm g)}{u^{0}c}\label{gemCSOm}\end{equation} in the weak Schwarzschild field characterized by Eq. (\ref{gemCSch}). Like the spin rotation (see Refs. \cite{Venema,PRD2007}), the OAM rotation can be observed in specially designed experiments. 
When the weak-field approximation is used in the corresponding uniformly accelerated frame ($\bm a=-\bm g$),  
$$\bm{\mathcal{E}}=-\frac{\bm au^{\widehat{0}}}{c}, \qquad \bm{\mathcal{B}}=0, \qquad
\bm{\Omega}_a=-\frac{u^{\widehat{0}}(\widehat{\bm\beta}\times\bm a)}{u^{0}c}.$$ Thus, the OAM rotates in this frame twice slower than in the weak Schwarzschild field (cf. Refs. \cite{PRDThomaspre,Lee,MashhoonObukhov}).

The obtained general result can find important applications in astrophysics which actively studies twisted photons and other particles \cite{Harwit,Tambu,Tamburini,Tamburi,photontwistd,orbitalHalleff}. In particular, our result describes the OAM dynamics in the Lense-Thirring metric being the weak-field approximation for the Kerr field. In this case \cite{PRDThomaspre},
\begin{equation}\begin{array}{c}
\bm{\mathcal{E}}=-\frac{Gm}{cr^3}\bm ru^{\widehat{0}}+\frac{3G}{c^2r^5}\left[{\bm r}
({\bm J}\cdot({\bm r}\times\widehat{\bm u})) -({\bm r}\times{\bm J})(\widehat{\bm u}\cdot{\bm r})\right], \\ \bm{\mathcal{B}}=-\frac{Gm}{cr^3}\bm r\times\widehat{\bm u}-\frac{G}{c^2r^3}\left [\frac{3(\bm r\cdot\bm J) \bm r}{r^2}-\bm J\right]u^{\widehat{0}}
\end{array}\label{gemLenT}\end{equation} and
\begin{equation}\begin{array}{c}
\bm{\Omega}=\frac{u^{\widehat{0}}}{u^{0}}\Biggl\{-\frac{2Gm}{cr^3}\Bigl(\widehat{\bm\beta}\times\bm r\Bigr)-\frac{3G}{c^2r^5}\biggl[(\widehat{\bm\beta}\times{\bm r})\Bigl({\bm J}\cdot(\widehat{\bm\beta}\times{\bm r})\Bigr)\\+\Bigl[\widehat{\bm\beta}\times({\bm r}\times{\bm J})\Bigr](\widehat{\bm\beta}\cdot{\bm r})\biggr] +\frac{G}{c^2r^3}\biggl[\frac{3(\bm r\cdot\bm J) \bm r}{r^2}-\bm J\biggr]\Biggr\}.
\end{array}\label{gemLenm}\end{equation}
A change of orbital polarization of twisted photons coming from binary stars can be rather significant due to strong gravitational fields of rotating invisible components (neutron stars or Kerr black holes). This change strongly depends on the distance between the photon trajectory and the invisible component. \emph{Exact} formulas for the gravitoelectromagnetic fields in arbitrary spacetimes 
have been derived in Refs. \cite{gravityDiracOST2,gravityDiracOST3}.

Equations (\ref{gelab}) and (\ref{final}) give also a general description of the orbital Hall effect in gravity. For twisted particles, it is similar to the gravitational spin Hall effect. The latter effect consists in the spin-dependent term, $\bm{\Omega}_s\cdot\bm s$, in the Hamiltonian, the dependence of the momentum on a spin polarization for a fixed energy (birefrigence of particle beams), 
and the gravitational spin-dependent Stern-Gerlach force, see Refs. \cite{MOancea,Oancealigh,Oancea2023} and references therein. The orbital Hall effect is defined by the term $\bm{\Omega}\cdot\widehat{\bm L}$ in the Hamiltonian (\ref{gelab}). The two Hall effects differ by the distinction between quantization laws of the spin and OAM resulting in multirefrigence of twisted particle beams with a fixed energy in general relativity. 

The spin-dependent term $\bm{\Omega}_s\cdot\bm s$ can violate the WEP \cite{Wang}. 
Existence of a similar violation caused by the OAM can be rigorously proven and described in the framework of \emph{classical} gravity. The velocity of any twisted particle has the \emph{OAM-dependent part} defined by the Hamilton equation $\widehat{\bm v}=\frac{d\widehat{\bm r}}{d\widehat{t}}=\frac{\partial \widehat{H}}{\partial\widehat{\bm p}}$. Let a twisted particle freely falls along the $z$ axis in the weak Schwarzschild field $\bm g=g\bm e_z$, has a nonzero initial momentum $\widehat{\bm p}{(0)}$, and its OAM be collinear to the $y$ axis ($\widehat{\bm L}=L_y\bm e_y$). As a rule, the initial momentum and OAM are collinear [$\widehat{\bm p}{(0)}=\widehat{p}_y{(0)}\bm e_y$]. As follows from Eqs. (\ref{gelab}) and (\ref{gemCSOm}), the WEP violation manifests itself in the OAM-dependent motion parallel or antiparallel to the $x$ axis (cf. Ref. \cite{Wang}).
The motion along the $z$ axis remains unchanged. In the weak-field approximation, $x^\mu\approx\widehat{x}^\mu$ and 
\begin{equation}
v_x=-\frac{2gL_y}{\epsilon},\qquad x(t)=-\frac{2gL_y}{\epsilon}t.
\label{violWEP}\end{equation} Evidently, the use of twisted photons with $\ell\gg1$ exceptionally simplifies the discovery of this effect, while spreading of twisted beams complicates its observation.

There is not any restriction on a mutual orientation of the intrinsic OAM and momentum of a twisted photon because such a photon is equivalent to a centroid with a nonzero effective mass \cite{photonPRA}. In contrast, the photon spin projection on the momentum direction is definite and the helicity of the photon is equal to $\mathfrak{h}=\widehat{\bm p}\cdot\bm s/(\hbar\widehat{p})=\pm1$.

As a rule, OAMs of twisted particles produced in laboratories are collinear to their momenta. In \emph{static} spacetimes, there is no orbital Hall effect for such particles. In these spacetimes, $\bm{\mathcal{B}}$ is orthogonal to $\widehat{\bm u}$ and the interaction Hamiltonian (\ref{gelab}) vanishes. In \emph{stationary} spacetimes, the gravitomagnetic field contains term which does not collinear to $\widehat{\bm u}$. Due to this term, $H$ is nonzero while the contribution from the gravitoelectric field vanishes. Thus, the gravitational orbital Hall effect exists and is observable for stationary spacetimes, e.g., the rotating frame and the Lense-Thirring and Kerr metrics. We should add that this effect takes also place for \emph{charged} beams in static spacetimes if the initial longitudinal polarization of beams is changed with electric and magnetic fields (see Ref. \cite{Manipulating}). After this procedure, the coframe vectors $\widehat{\bm L}$ and $\widehat{\bm u}$ become noncollinear and the gravitational orbital Hall effect can be observed. 

Importantly, the gravitational orbital Hall effect can be much stronger than the corresponding spin effect because the OAM $\widehat{L}=\hbar\ell$ can be rather large. The general form of the gravitational OAM-dependent Stern-Gerlach force reads 
$$\bm F_{SG}^{(OAM)}=-\nabla H=-\nabla(\bm{\Omega}\cdot\widehat{\bm L}).$$ This force can be rather significant near rotating astronomical objects. 

The multirefrigence of beams is also caused by the interaction Hamiltonian $H$ and is defined by (see Supplemental Material for details)
$$\delta k=\frac{\Omega\widehat{L}_\Omega}{\hbar v_{(0)}\cos{\theta}},\qquad \cos{\theta}=\frac{\bm v_{(0)}\cdot\bm p_{(0)}}{v_{(0)} p_{(0)}},$$
where $\widehat{L}_\Omega=-\widehat{L},-\widehat{L}+1,\dots, 0,\dots,\widehat{L}$ and $\theta$ is the angle between the vectors $\bm v_{(0)}$ and $\bm p_{(0)}$.

In summary, the general solutions of problems of the rotation of intrinsic OAM and the orbital Hall effect for twisted particles in arbitrary gravitational fields have been obtained. We used the similarity between the OAM and spin and considered the OAM motion in different LLFs. Amazingly, our approach does not need Maxwell equations in curved spacetimes for a description of twisted photons.
We have derived the exact general equation (\ref{final}) rigorously defining the OAM dynamics in any Riemannian spacetimes. The general description of different manifestations of the orbital Hall effect for any twisted particles has also been presented. Our short analysis shows that the results obtained can find important practical applications.

\begin{acknowledgments}
This work was supported by the Russian Science Foundation (Grant No. 25-72-30005). \end{acknowledgments}

\section{Supplemental Material}

\subsection{Comparison of dynamics of orbital angular momentum for massive ultrarelativistic and massless beams}

The determination of the difference between the dynamics of intrinsic orbital angular momentum (OAM) in Riemannian spacetimes for massive and massless beams is important for twisted photons. Such photons are always ultrarelativistic and have nonzero effective masses \cite{photonPRA}. For such particles, the OAM rotation is defined by the general equation (24) in the main text. However, it is instructive to determine the uncertainties caused by the use of the corresponding equation for massless photon beams. In this case, one considers plane waves of photons instead of photon beams and introduces the fixed orbital helicity of photons $\mathfrak{h}^{(OAM)}=\bm p\cdot\bm L/(\hbar p)=\pm\ell$. In gravity, fundamental properties of photons, e.g., the orthogonality of the electric and magnetic field and the momentum, manifest themselves in local Lorentz frames. In such frames, the orbital helicity and the OAM of photons are \emph{approximately} given by $\mathfrak{h}^{(OAM)}=\widehat{\bm p}\cdot\widehat{\bm L}/(\hbar\widehat{p})=\pm\ell$ and $\widehat{\bm L}=\hbar\ell\widehat{\bm p}/\widehat{p}=\hbar\ell\widehat{\bm u}/|\widehat{\bm u}|$. In this case, the OAM rotates with the same angular velocity as the unit vector $\bm N=\widehat{\bm u}/|\widehat{\bm u}|$. The rotation of this vector is defined by
$$\frac{d\bm N}{dt}=\frac{1}{|\widehat{\bm u}|^3}\left(\widehat{\bm u}\times\left(\frac{d\widehat{\bm u}}{dt}\times\widehat{\bm u}\right)\right).
$$
As follows from this equation and Eq. (8) in the main text,
\begin{equation}
\begin{array}{c} \frac{d\widehat{\bm{u}}}{dt}=\frac{u^{\widehat{0}}\bm{\mathcal{E}}+
\widehat{{\bm
u}}\times\bm{\mathcal{B}}}{u^0},\qquad \frac{d\bm N}{dt}=\frac{\widehat{{\bm
u}}\times\bm{\mathcal{B}}}{u^0|\widehat{\bm u}|}-\frac{u^{\widehat{0}}(\widehat{{\bm
u}}\times(\widehat{{\bm
u}}\times\bm{\mathcal{E}}))}{u^0|\widehat{\bm u}|^3}.
\end{array} \label{forceft} 
\end{equation} Since $\widehat{\bm u}=u^{\widehat{0}}\widehat{\bm\beta}$ and $|\widehat{\bm u}|^2=(u^{\widehat{0}})^2-1$, we obtain
\begin{equation}
\begin{array}{c} \frac{d\bm N}{dt}=\bm\omega\times\bm N,\qquad \bm\omega=-\frac{\bm{\mathcal{B}}}{u^0}+\frac{(u^{\widehat{0}})^2(\widehat{\bm\beta}\times\bm{\mathcal{E}})}{u^0\bigl[(u^{\widehat{0}})^2-1\bigr]}.
\end{array} \label{appav} 
\end{equation}
As a result, we find the following difference between the exact and approximate angular velocities of OAM rotation in Riemannian spacetimes:
\begin{equation}
\begin{array}{c} \bm\Omega-\bm\omega=-\frac{\widehat{\bm\beta}\times\bm{\mathcal{E}}}{u^0\bigl[(u^{\widehat{0}})^2-1\bigr]}.
\end{array} \label{appav} 
\end{equation}

In terrestrial conditions, the ratio $|\bm\Omega-\bm\omega|/\Omega$ is of the order of $\bigl(u^{\widehat{0}}\bigr)^{-2}=\gamma^{-2}$, where $\gamma$ is the Lorentz factor. For twisted photons, $(1-\beta)\sim\gamma^{-2}\sim10^{-5}-10^{-4}$ \cite{Giovannini,AlfanoNolanBessel,Bouchard,Lyons} and, as a rule, the approximation $\widehat{\bm L}=\hbar\ell\widehat{\bm p}/\widehat{p}=\hbar\ell\widehat{\bm u}/|\widehat{\bm u}|$ is admittable for a description of the OAM rotation. This approximation is inappropriate for the analysis of multirefrigence of twisted beams because it predicts that the OAM has only two projections $\pm\widehat{L}$ on the momentum
direction. Since a photon centroid is effectively massive, it is characterized by the rest frame and $2\ell + 1$ OAM projections on a chosen direction.

\subsection{Multirefrigence of twisted beams}

In this subsection, we consider the multirefrigence of twisted beams. For this purpose, we need to pass to the world frame. Since $H$ expressed by Eq. (23) in the main text is small compared to the total Hamiltonian ${\cal H}$, there is no need to express $\widehat{\bm{v}}$ in terms of world frame quantities. However, we should note that $$v^{\widehat{i}}=\frac{dx^{\widehat{i}}}{dx^{\widehat{0}}}=\frac{e^{\widehat{i}}_\mu dx^\mu}{e^{\widehat{0}}_\nu dx^\nu}=\frac{e^{\widehat{i}}_0 dx^0+e^{\widehat{i}}_j dx^j}{e^{\widehat{0}}_0 dx^0}=\frac{e^{\widehat{i}}_0}{e^{\widehat{0}}_0}+\frac{e^{\widehat{i}}_j}{e^{\widehat{0}}_0}v^j,$$ where $v^i=-\frac{\partial{\cal H}}{\partial p_i}$ is the world velocity and the Schwinger tetrad is used.

For particles without spin and intrinsic orbital angular momentum (OAM), the Hamiltonian has the form \cite{Cogn}
\begin{equation}
{\cal H}={\cal H}_{(0)}= \left(\frac{m^2 - G^{ij}p_ip_j} {g^{00}}\right)^{1/2} -
\frac{g^{0i}p_i}{{g}^{00}}, \qquad G^{ij}=g^{ij}-\frac{g^{0i}g^{0j}}{g^{00}}. \label{clCog}
\end{equation} For a fixed energy (${\cal H}_{(0)}=E$), the canonical momentum $\bm P_{(0)}=-(p_{(0)i})$ is also fixed, but depends on the direction of the beam. Importantly, the canonical momentum $\bm P=-(p_i)$ differs from the kinetic momentum $\bm p=(mu^i)$ used in the main text. 
The total Hamiltonian of a twisted particle is given by 
\begin{equation}
{\cal H}={\cal H}_{(0)}+H={\cal H}_{(0)}+\bm{\Omega}\cdot\widehat{\bm L}={\cal H}_{(0)}+\Omega \widehat{L}_\Omega,
\label{clHH}
\end{equation} where $\bm{\Omega}$ is the angular velocity of the OAM rotation and $\widehat{L}_\Omega=-\widehat{L},-\widehat{L}+1,\dots, 0,\dots,\widehat{L}$.

In stationary states, ${\cal H}=E=const$ and the canonical momentum $\bm P$ depends on $\Omega \widehat{L}_\Omega$. Equation (\ref{clCog}) defines the following relation between small changes of ${\cal H}_{(0)}$ and $\bm P$:
\begin{equation} \begin{array}{c}
\delta{\cal H}_{(0)}= \frac{\partial{\cal H}_{(0)}}{\partial p_i}\biggl|_{\bm P=\bm P_{(0)}}\delta p_i=-v_{(0)}^i\delta p_i=\bm v_{(0)}\cdot\delta\bm P,\\  v_{(0)}^i=G^{ij}p_{(0)j}\left(\frac{m^2 - G^{ij}p_{(0)i}p_{(0)j}} {g^{00}}\right)^{-1/2} +
\frac{g^{0i}}{{g}^{00}}, \label{clCogde}
\end{array} \end{equation} where $\bm v_{(0)}$ is the world velocity for a particle without the spin and the intrinsic OAM. 

The beam can be characterized by any possible value of $\widehat{L}_\Omega$ and, in the general case, it is a superposition of partial beams with the same $\widehat{L}$ and different $\widehat{L}_\Omega$. Partial beams interfere when they have the same energy $E$ and different canonical momenta $\bm P=\bm P_{(0)}+\delta\bm P$, where the vectors $\bm P$ and $\delta\bm P$ are collinear. In this case, $\Omega\widehat{L}_\Omega=\bm v_{(0)}\cdot\delta\bm P$ and 
$$\delta P=\hbar\, \delta k=\frac{\Omega\widehat{L}_\Omega}{ v_{(0)}\cos{\theta}},\qquad \cos{\theta}=\frac{\bm v_{(0)}\cdot\bm P_{(0)}}{v_{(0)} P_{(0)}},$$ where $\theta$ is the angle between the vectors $\bm v_{(0)}$ and $\bm P_{(0)}$.

The quantity $\delta k$ defines the multirefrigence of twisted beams. 

One can introduce $2\widehat{L}+1$ basic wave functions which describe partial beams with different integer  $\widehat{L}_\Omega$ and the corresponding $k=P_{(0)}/\hbar+\delta k$. Any wave function defining a twisted beam with an arbitrary OAM direction can be presented as a coherent superposition of these basic wave functions. Like the spin rotation, the OAM rotation can be equivalently presented as a result of evolution of the latter functions. 

\end{document}